\documentclass[superscriptaddress]{revtex4}
\usepackage{amssymb}
\usepackage[latin9]{inputenc}
\usepackage{graphicx}

\makeatletter \@ifundefined{textcolor}{} {
 \definecolor{BLACK}{gray}{0}
 \definecolor{WHITE}{gray}{1}
 \definecolor{RED}{rgb}{1,0,0}
 \definecolor{GREEN}{rgb}{0,1,0}
 \definecolor{BLUE}{rgb}{0,0,1}
 \definecolor{CYAN}{cmyk}{1,0,0,0}
 \definecolor{MAGENTA}{cmyk}{0,1,0,0}
 \definecolor{YELLOW}{cmyk}{0,0,1,0}
 }
\makeatother

\begin{document}

\title{How Perfect a Gluon Plasma Can Be in Perturbative QCD?}
\author{Jiunn-Wei Chen}
\affiliation{Department of Physics and Center for Theoretical Sciences, National Taiwan
University, Taipei 10617, Taiwan}
\author{Jian Deng}
\affiliation{School of Physics, Shandong University, Shandong 250100, People's Republic
of China}
\author{Hui Dong}
\affiliation{School of Physics, Shandong University, Shandong 250100, People's Republic
of China}
\author{Qun Wang}
\affiliation{Interdisciplinary Center for Theoretical Study and Department of Modern
Physics, University of Science and Technology of China, Anhui 230026,
People's Republic of China}

\begin{abstract}
The shear viscosity to entropy density ratio, $\eta /s$, characterizes how
perfect a fluid is. We calculate the leading order $\eta /s$ of a gluon
plasma in perturbation using the kinetic theory. The leading order
contribution only involves the elastic $gg\leftrightarrow gg$ (22) process
and the inelastic $gg\leftrightarrow ggg$ (23) process. The
Hard-Thermal-Loop (HTL) treatment is used for the 22 matrix element, while
the exact matrix element in vacuum is supplemented by the gluon Debye mass
insertion for the 23 process. Also, the asymptotic mass is used for the
external gluons in the kinetic theory. The errors from not implementing HTL
and the Landau-Pomeranchuk-Migdal effect in the 23 process, and from the
uncalculated higher order corrections, are estimated. Our result for $\eta /s
$ lies between that of Arnold, Moore and Yaffe (AMY) and Xu and Greiner
(XG). Our result shows that although the finite angle contributions are
important at intermediate $\alpha _{s}$\ ($\alpha _{s}\sim 0.01$-$0.1$), the
22 process is still more important than 23 when $\alpha _{s}\lesssim 0.1$.
This is in qualitative agreement with AMY's result. We find no indication
that the proposed perfect fluid limit $\eta /s\simeq 1/(4\pi )$\ can be
achieved by perturbative QCD alone.
\end{abstract}

\maketitle

\section{Introduction}

A \textit{perfect fluid} is a system with zero shear and bulk viscosities, $%
\eta $ and $\zeta $, and no dissipation. These conditions can be satisfied
for a superfluid at zero temperature where only the super fluid component
exists, but a sharper description is with the dimensionless ratios $\eta /s$
and $\zeta /s$, where $s$, the entropy density, vanishes for the superfluid
component as well.\ While $\zeta /s$ can still be zero for scaling invariant
systems, the situation for $\eta /s$ is more subtle.

In general, stronger interactions implies a smaller $\eta $. Thus, a perfect
fluid with the smallest $\eta /s$ is likely to be strongly interacting which
requires non-perturbative tools to compute it. The anti-de-Sitter
space/conformal field theory correspondence (AdS/CFT) \cite%
{Maldacena:1997re,Gubser:1998bc,Witten:1998qj} allows the $\eta /s$\ of
strongly interacting CFT's to be computed in weakly interaction
gravitational theories. A universal number $\eta /s=1/(4\pi )$ is found for
every CFT with a gravity dual in the large $N$, with $N$ the size of the
gauge group, and infinite t'Hooft coupling limit \cite%
{Buchel:2003tz,Kovtun:2004de,Buchel:2004qq}. With this result, together with
the connection to the uncertainty principle through the relation $\eta
/s\sim \Delta E\Delta t$, with $\Delta E$ and $\Delta t$ the mean energy and
life time of quasiparticles, Kovtun, Son, and Starinets \ (KSS) \cite%
{Kovtun:2004de} conjectured that the strongly interacting CFT value $1/(4\pi
)$ is the minimum bound for $\eta /s$\ for all physical systems.

Theoretically, there are several attempts to evade this bound. It is found
that $\eta /s$ can be as small as possible (but still non-negative) in a
carefully engineered meson system \cite{Cohen:2007qr,Cherman:2007fj},
although the system is meta-stable. Also, in strongly interacting CFTs, $1/N$
corrections can be negative \cite{Kats:2007mq,Brigante:2007nu}\ and can
modify the $\eta /s$ bound slightly \cite{Brigante:2008gz,Buchel:2008vz}.

Experimentally, there are intensive interests to find the most perfect fluid
(see \cite{Kapusta:2008vb,Schafer:2009dj} for recent reviews). The smallest $%
\eta /s$ known so far is realized in a system of hot and dense matter
thought to be quark gluon plasma just above the phase transition temperature
produced at RHIC \cite{Arsene:2004fa,Back:2004je,Adcox:2004mh} with $\eta
/s=0.1\pm 0.1(\mathrm{theory})\pm 0.08(\mathrm{experiment})$ \cite%
{Luzum:2008cw}. A robust upper limit $\eta /s<5\times 1/(4\pi )$ was
extracted by another group \cite{Song:2008hj} and a lattice computation of
gluon plasma yields $\eta /s=0.134(33)$ \cite{etas-gluon-lat}. Progress has
been made in cold unitary fermi gases as well. An analysis of the damping of
collective oscillations gives $\eta /s\gtrsim 0.5$ \cite{Schafer,Turlapov}.
Even smaller values of $\eta /s$ are indicated by recent data on the
expansion of rotating clouds \cite{Clancy,Thomas} but more careful analyses
are needed \cite{Schaefer2,Schaefer3}.

Even if the $1/(4\pi )$ bound for $\eta /s$ turns out to be invalid, it is
still interesting to use it as a bench mark value for the perfectness of
fluids. It was found that based on the perturbative QCD (PQCD) analysis of
Arnold, Moore and Yaffe (AMY) \cite{Arnold:2000dr,Arnold:2003zc}, the
measured $\eta /s$ at RHIC cannot be explained by PQCD (for a recent review,
see, e.g., \cite{Arnold:2007pg}). This strongly interacting QGP picture is
very different from the conventional picture of weakly interacting QGP\ and
is considered as one of the most surprising discoveries at the RHIC.

However, a recent perturbative QCD calculation\ of $\eta /s$ of a gluon
plasma by Xu and Greiner (XG) \cite{Xu:2007ns} shows that the dominant
contribution comes from the inelastic $gg\leftrightarrow ggg$ (23) process
instead of the elastic $gg\rightarrow gg$\ (22) process. In particular, the
23 process is 7 times more important than 22. Thus, $\eta /s\simeq 1/4\pi $
can be achieved when the strong coupling constant $\alpha _{s}\simeq 0.6$.
In Ref. \cite{Wesp:2011yy}, XG and their collaborators improve their calculation 
using the Kubo relation and give smaller contribution from the 23 process: 
about 5 times (2$\sim$9 times) the 22 process.     
Thus, the conventional weakly interacting QGP could still be valid. This is
in sharp contrast to AMY's result where the 23 process only gives $\sim 10\%$
correction to the 22 process.

Both XG and AMY\ use kinetic theory for their calculations. The main
differences are (i) XG uses a parton cascade model \cite{Xu:2004mz} to solve
the Boltzmann equation and, for technical reasons, gluons are treated as a
classical gas instead of a bosonic gas. On the other hand, AMY solves the
Boltzmann equation for a bosonic gas. (ii) AMY approximates the $%
Ng\leftrightarrow (N+1)g$ processes, $N=2,3,4\ldots $, by the $%
g\leftrightarrow gg$ splitting in the collinear limit where the two gluon
splitting angle is higher order. XG uses the soft gluon bremsstrahlung limit
where one of the gluon momenta in the final state of $gg\rightarrow ggg$ is
soft but it can have a large splitting angle with its mother gluon.

In an earlier attempt to resolve the discrepancy between XG's and AMY's
results \cite{Chen:2009sm}, a Boltzmann equation computation of $\eta $ is
carried out without taking the classical gluon approximation (like AMY's
approach) but the soft gluon bremsstrahlung limit is applied to the 23
matrix element (like XG's approach, modulo a factor 2 in the 23 matrix
element squared; see \cite{Chen:2009sm} for details). It was found that the
classical gas approximation does not cause a significant error in $\eta /s$
(although the individual errors on $\eta $ and $s$ are larger). However, the
result is sensitive to whether the soft gluon bremsstrahlung limit is
imposed on the phase space or not. If this limit is imposed, the result is
closer to AMY's; if not, the result is closer to XG's. This raises the
concern whether this approximation is good for computing $\eta $.

The goal of this paper is to settle this issue by removing both the soft
gluon bremsstrahlung approximation and the collinear approximation to the 23
process. The leading order [$O(\alpha _{s}^{-2})$] contribution to $\eta $
only involves the 22 and 23 processes \cite{Arnold:2003zc} (the power
counting for 22, 23 and other processes are reproduced in \cite{Chen:2009sm}%
). In this paper, the Hard-Thermal-Loop (HTL) treatment is used for the 22
matrix element, while the exact matrix element in vacuum is supplemented by
the gluon Debye mass insertion for the 23 process. Also, the Debye mass is
used for the external gluon mass in the kinetic theory as well. The errors
from not implementing HTL and the Landau-Pomeranchuk-Migdal effect in the 23
process, and from the uncalculated higher order corrections, are also
estimated.

\section{Kinetic theory beyond the soft or collinear gluon approximations}

Using the Kubo formula, $\eta $ can be calculated through the linearized
response function of a thermal equilibrium state 
\begin{equation}
\eta =-\frac{1}{5}\int_{-\infty }^{0}\mathrm{d}t^{\prime }\int_{-\infty
}^{t^{\prime }}\mathrm{d}t\int \mathrm{d}x^{3}\langle \left[
T^{ij}(0),T^{ij}(\mathbf{x},t)\right] \rangle ,
\end{equation}%
where $T^{ij}$ is the spatial part of the off-diagonal energy momentum
tensor. In a leading order (LO) expansion of the coupling constant, there
are an infinite number of diagrams \cite{Jeon:1994if,Jeon:1995zm}. However,
it is proven that the summation of the LO diagrams in a weakly coupled $\phi
^{4}$ theory \cite%
{Jeon:1994if,Jeon:1995zm,Carrington:1999bw,Wang:1999gv,Hidaka:2010gh} or in
hot QED \cite{Gagnon:2007qt} is equivalent to solving the linearized
Boltzmann equation with temperature-dependent particle masses and scattering
amplitudes. The conclusion is expected to hold in weakly coupled systems and
can as well be used to compute the LO transport coefficients in QCD-like
theories \cite{Arnold:2000dr,Arnold:2003zc}, hadronic gases \cite%
{Prakash:1993bt,Dobado:2003wr,Dobado:2001jf,Chen:2006iga,Chen:2007xe,Itakura:2007mx}
and weakly coupled scalar field theories \cite%
{Jeon:1994if,Jeon:1995zm,Carrington:1999bw,Wang:1999gv,Moore:2007ib,Chen:2007jq}%
.

The Boltzmann equation of a hot gluon plasma describes the evolution of the
color and spin averaged gluon distribution function $f_{p}(x)$ which is a
function of space-time $x=(t,\mathbf{x})$ and momentum $p=(E_{p},\mathbf{p})$%
. The infinitesimal deviation of $f_{p}(x)$ from its equilibrium value $%
f_{p}^{eq}=(e^{v\cdot p/T}-1)^{-1}$ is denoted as 
\begin{equation}
f_{p}=f_{p}^{eq}[1-\chi _{p}(1+f_{p}^{eq})],
\end{equation}%
where $\chi _{p}\equiv \chi (x,p)$ can be parametrized as 
\begin{equation}
\chi _{p}=\frac{A(p)}{T}\nabla \cdot \mathbf{v}+\frac{B_{ij}(p)}{T}\frac{1}{2%
}\left( \frac{\partial v^{i}}{\partial x^{j}}+\frac{\partial v^{j}}{\partial
x^{i}}-\frac{2}{3}\delta _{ij}\nabla \cdot \mathbf{v}\right) ,
\label{eq:chi}
\end{equation}%
at the leading order of the derivative expansion of the fluid velocity $%
v(x)=(v^{0},v)$. $T=T(x)$ is the local temperature, $\mathbf{\hat{p}}$ is
the unit vector in the $\mathbf{p}$ direction. $B_{ij}(p)=B(p)(\mathbf{\hat{p%
}}^{i}\mathbf{\hat{p}}^{j}-\frac{1}{3}\delta _{ij})$. $A(p)$ and $B(p)$ are
functions of $\mathbf{p}$ which will be fixed by the Boltzmann equation
corresponding to the bulk and shear viscosities, respectively. In this work
we will just focus on the shear viscosity calculation.

The Boltzmann equation \cite%
{Heinz:1984yq,Elze:1986qd,Biro:1993qt,Blaizot:1999xk,Baier:2000sb,Wang:2001dm}
for the gluon plasma reads 
\begin{eqnarray}
\frac{p^{\mu }}{E_{p}}\partial _{\mu }f_{p} &=&\frac{1}{N_{g}}\sum_{(n,l)}%
\frac{1}{N(n,l)}\int_{1\cdots \left( n-1\right) }d\Gamma _{1\cdots
l\rightarrow (l+1)\cdots (n-1)p}  \nonumber \\
&&\times \left[ (1+f_{p})\prod_{r=1}^{l}f_{r}%
\prod_{s=l+1}^{n-1}(1+f_{s})-f_{p}\prod_{r=1}^{l}(1+f_{r})%
\prod_{s=l+1}^{n-1}f_{s}\right] ,  \label{eq:be1}
\end{eqnarray}%
where the collision rates are given by 
\begin{equation}
d\Gamma _{1\cdots l\rightarrow (l+1)\cdots (n-1)p}\equiv \prod_{j=1}^{n-1}%
\frac{d^{3}\mathbf{p}_{j}}{(2\pi )^{3}2E_{j}}\frac{1}{2E_{p}}|M_{1\cdots
l\rightarrow (l+1)\cdots (n-1)p}|^{2}(2\pi )^{4}\delta
^{4}(\sum_{r=1}^{l}p_{r}-\sum_{s=l+1}^{n-1}p_{s}-p).  \label{eq:rate}
\end{equation}%
$N_{g}=16$ is the color and spin degeneracy of a gluon. The $i$-th gluon is
labeled as $i$ while the $n$-th gluon is labeled as $p$. For a process with $%
l$ initial and $(n-l)$ final gluons, the symmetry factor $N(n,l)=l!(n-l-1)!$%
. For example, processes $12\rightarrow 3p$, $12\rightarrow 34p$, $%
123\rightarrow 4p$ yield $(n,l)=(4,2),(5,2),(5,3)$ and $N(n,l)=2,4,6$,
respectively.

In vacuum, the matrix element squared for the 22 process is 
\begin{equation}
|M_{12\rightarrow 34}|^{2}=\frac{9}{2}(4\pi )^{2}N_{g}^{2}\alpha
_{s}^{2}\left( 3-\frac{tu}{s^{2}}-\frac{su}{t^{2}}-\frac{st}{u^{2}}\right) ,
\label{eq:matrix-e22}
\end{equation}%
where $\alpha _{s}=g^{2}/(4\pi )$ is the strong coupling constant, and $%
(s,t,u)$ are the Mandelstam variables $s=(p_{1}+p_{2})^{2}$, $%
t=(p_{1}-p_{3})^{2}$ and $u=(p_{1}-p_{4})^{2}$.

For the 23 process \cite{Ellis:1985er,Gottschalk:1979wq}, under the
convention $\sum_{i=1}^{5}p_{i}=0$, we have 
\begin{eqnarray}
\left\vert M_{12345\rightarrow 0}\right\vert ^{2} &=&\left\vert
M_{0\rightarrow 12345}\right\vert ^{2}  \nonumber \\
&=&54\pi ^{3}N_{g}^{2}\alpha _{s}^{3}\left[ \left( 12\right) ^{4}+\left(
13\right) ^{4}+\left( 14\right) ^{4}+\left( 15\right) ^{4}+\left( 23\right)
^{4}\right.  \nonumber \\
&&\left. +\left( 24\right) ^{4}+\left( 25\right) ^{4}+\left( 34\right)
^{4}+\left( 35\right) ^{4}+\left( 45\right) ^{4}\right]  \nonumber \\
&&\times \sum\limits_{\mathrm{perm}\left\{ 1,2,3,4\right\} }\frac{1}{\left(
12\right) \left( 23\right) \left( 34\right) \left( 45\right) \left(
51\right) },  \label{eq:m3}
\end{eqnarray}
where $(ij)\equiv p_{i}\cdot p_{j}$ and the sum is over all permutations of $%
\{1,2,3,4\}$. To convert to the convention $p_{1}+p_{2}=p_{3}+p_{4}+p_{5}$,
we just perform the replacement: 
\begin{eqnarray}
\left\vert M_{12\rightarrow 345}\right\vert ^{2} &=&\left. \left\vert
M_{0\rightarrow 12345}\right\vert ^{2}\right\vert _{p_{1}\rightarrow
-p_{1},p_{2}\rightarrow -p_{2}},  \nonumber \\
\left\vert M_{345\rightarrow 12}\right\vert ^{2} &=&\left. \left\vert
M_{12345\rightarrow 0}\right\vert ^{2}\right\vert _{p_{1}\rightarrow
-p_{1},p_{2}\rightarrow -p_{2}}.  \label{m4}
\end{eqnarray}

In the medium, the gluon thermal mass effect serves as the infrared (IR)
cut-off to regularize IR sensitive observables. The most singular part of
Eq.(\ref{eq:matrix-e22}) comes from the collinear region (i.e. either $%
t\approx 0$ or $u\approx 0$) which can be regularized by the
Hard-Thermal-Loop (HTL) corrections to the gluon propagators \cite%
{Weldon:1982aq,Pisarski:1988vd} and yields \cite{Heiselberg:1996xg}, 
\begin{equation}
|M_{12\rightarrow 34}|^{2}\approx \frac{1}{4}(12\pi \alpha
_{s})^{2}N_{g}^{2}(4E_{1}E_{2})^{2}\left\vert \frac{1}{\mathbf{q}^{2}+\Pi
_{L}}-\frac{(1-\overline{x}^{2})\cos \phi }{\mathbf{q}^{2}(1-\overline{x}%
^{2})+\Pi _{T}}\right\vert ^{2},  \label{eq:htl}
\end{equation}%
where $q=p_{1}-p_{3}=(q_{0},\mathbf{q}),$ $\overline{x}=q_{0}/|\mathbf{q}|$
and $\phi $ is the angle between $\hat{\mathbf{p}}_{1}\times \hat{\mathbf{q}}
$ and $\hat{\mathbf{p}}_{2}\times \hat{\mathbf{q}}$. The HTL self-energies $%
\Pi _{L}$ (longitudinal) and $\Pi _{T}$ (transverse) are given by 
\begin{eqnarray}
\Pi _{L} &=&m_{D}^{2}\left[ 1-\frac{\overline{x}}{2}\ln \frac{1+\overline{x}%
}{1-\overline{x}}+i\frac{\pi }{2}\overline{x}\right] ,  \nonumber \\
\Pi _{T} &=&m_{D}^{2}\left[ \frac{\overline{x}^{2}}{2}+\frac{\overline{x}}{4}%
(1-\overline{x}^{2})\ln \frac{1+\overline{x}}{1-\overline{x}}-i\frac{\pi }{4}%
\overline{x}(1-\overline{x}^{2})\right] .  \label{HTL}
\end{eqnarray}%
The external gluon mass $m_{\infty }$\ (i.e. the asymptotic mass) is the
mass for an on-shell transverse gluon, and $m_{\infty }^{2}=\Pi _{T}\left(
\left\vert \overline{x}\right\vert =1\right) =m_{D}^{2}/2$\ both in the HTL
approximation and in the full one-loop result.

Previous investigations of the thermodynamics within resummed perturbation
theory showed that the most important plasma effects are the thermal masses $%
\sim gT$\ acquired by the hard thermal particles \cite%
{Blaizot:2000fc,Andersen:2002ey,CaronHuot:2007gq}. So a simpler (though less
accurate) treatment for the regulator is to insert the Debye mass $%
m_{D}=(4\pi \alpha _{s})^{1/2}T$ to the gluon propagator such that in the
center-of-mass (CM) frame, 
\begin{equation}
|M_{12\rightarrow 34}|_{CM}^{2}\approx (12\pi \alpha _{s})^{2}N_{g}^{2}\frac{%
s^{2}}{(\mathbf{q}_{T}^{2}+m_{D}^{2})^{2}},  \label{MD}
\end{equation}%
where $\mathbf{q}_{T}$ is the transverse component of $\mathbf{q}$ with
respect to $\mathbf{p}_{1}$. It can be shown easily that Eqs. (\ref{eq:htl})
and (\ref{MD}) coincide in the CM frame in vacuum. This treatment was used
in Refs. \cite{Xu:2007ns,Biro:1993qt,Chen:2009sm}. We will show both results
using HTL and $m_{D}$ for the mass of the internal gluon propagators for
comparison.

For the 23 process, because the matrix element is already quite complicated,
we will just take $m_{D}$ as the regulator for internal gluons and estimate
the errors. In the $\sum_{i=1}^{5}p_{i}=0$ convention, one can easily show
that an internal gluon will have a momentum of $\pm (p_{i}+p_{j})$ rather
than $\pm (p_{i}-p_{j})$. Therefore, the gluon propagator factors $(ij)$ in
the denominator of Eq. (\ref{eq:m3}), should be modified to 
\begin{eqnarray}
(ij) &=&\frac{1}{2}[(p_{i}+p_{j})^{2}-m_{D}^{2}]  \nonumber \\
&=&p_{i}\cdot p_{j}\ +\frac{2m_{g}^{2}-m_{D}^{2}}{2},  \label{(ij)}
\end{eqnarray}%
where we use $m_{g}$ to denote the external gluon mass. Then one applies Eq.
(\ref{m4}) for the Boltzmann equation. In the numerator, the $(ij)^{4}$
combination is set by $T$ and is $\mathcal{O}(T^{8})$. So we can still apply
the substitution of Eq.(\ref{(ij)}), even if the $(ij)$ factors might not
have the inverse propagator form. The error is $\sim m_{D}^{2}(ij)^{3}=%
\mathcal{O}(\alpha _{s}T^{8})$, which is higher order in $\alpha _{s}$.

It is instructive to show that Eqs. (\ref{eq:m3},\ref{m4}) and (\ref{(ij)})
give the correct soft bremsstrahlung limit. Using the light-cone variable 
\begin{eqnarray}
p &=&\left( p^{+},p^{-},\mathbf{p}_{\perp }\right)  \nonumber \\
&\equiv &\left( p_{0}+p_{3},p_{0}-p_{3},p_{1},p_{2}\right) ,
\end{eqnarray}%
we can rewrite one momentum configuration in the CM frame in terms of $%
p,p^{\prime },q$ and $k$\textbf{: }$p_{1}=p$, $p_{2}=p^{\prime }$, $%
p_{3}=p+q-k$, $p_{4}=p^{\prime }-q$ and $p_{5}=k$, with 
\begin{eqnarray}
p &=&\left( \sqrt{s},m_{g}^{2}/\sqrt{s},0,0\right) ,  \nonumber \\
p^{\prime } &=&\left( m_{g}^{2}/\sqrt{s},\sqrt{s},0,0\right) ,  \nonumber \\
k &=&\left( y\sqrt{s},\left( k_{\perp }^{2}+m_{g}^{2}\right) /y\sqrt{s}%
,k_{\perp },0\right) ,  \nonumber \\
q &=&\left( q^{+},q^{-},q_{\perp }\right) .
\end{eqnarray}%
The on-shell condition $p_{3}^{2}=p_{4}^{2}=m_{g}^{2}$ yields 
\begin{eqnarray}
q^{+} &\simeq &-q_{\perp }^{2}/\sqrt{s},  \nonumber \\
q^{-} &\simeq &\frac{k_{\perp }^{2}+yq_{\perp }^{2}-2y\mathbf{k}_{\perp
}\cdot \mathbf{q}_{\perp }+(1-y+y^{2})m_{g}^{2}}{y\left( 1-y\right) \sqrt{s}}%
.
\end{eqnarray}%
Taking the large $s$ limit, then the $y\rightarrow 0$, we obtain 
\begin{equation}
\left\vert M_{12\rightarrow 345}\right\vert _{CM}^{2}=\sum\limits_{\mathrm{%
perm}\left\{ 3,4,5\right\} }3456\pi ^{3}N_{g}^{2}\alpha _{s}^{3}\frac{s^{2}}{%
\left( k_{\perp }^{2}+m_{g}^{2}\right) \left( q_{\perp
}^{2}+m_{D}^{2}\right) \left[ \left( \mathbf{k}_{\perp }-\mathbf{q}_{\perp
}\right) ^{2}+m_{D}^{2}\right] },  \label{reduce-GB}
\end{equation}%
where the permutation is over all final state gluon configurations. We see
that Eq. (\ref{reduce-GB}) reduces to the Gunion-Bertsch formula \cite%
{Gunion:1981qs} after taking $m_{D},m_{g}\rightarrow 0$. A similar
derivation of can be found in Ref. \cite{Das:2010hs,Abir:2010kc}.

\section{Beyond variation -- solving for $\protect\eta $ systematically}

Following the standard procedure, the shear viscosity is related to $B(p)$
by, 
\begin{equation}
\eta =\frac{N_{g}}{10T}\int \frac{d^{3}p}{(2\pi )^{3}E_{p}}
f_{p}^{eq}(1+f_{p}^{eq})\left( \mathbf{p}^{i}\mathbf{p}^{j}-\frac{1}{3}
\delta _{ij}\mathbf{p}^{2}\right) B_{ij}(p).  \label{eq:eta}
\end{equation}
$B_{ij}(p)$ satisfies the constraint derived from the linearized Boltzmann
equation, 
\begin{eqnarray}
\mathbf{p}^{i}\mathbf{p}^{j}-\frac{1}{3}\delta _{ij}\mathbf{p}^{2} &=&\frac{%
E_{p}}{2N_{g}}\int_{123}d\Gamma _{12\rightarrow
3p}f_{1}^{eq}f_{2}^{eq}(1+f_{3}^{eq})(f_{p}^{eq})^{-1}  \nonumber \\
&&\times \lbrack B_{ij}(p)+B_{ij}(p_{3})-B_{ij}(p_{1})-B_{ij}(p_{2})] 
\nonumber \\
&&+\frac{E_{p}}{4N_{g}}\int_{1234}d\Gamma
_{12;34p}(1+f_{1}^{eq})(1+f_{2}^{eq})f_{3}^{eq}f_{4}^{eq}(1+f_{p}^{eq})^{-1}
\nonumber \\
&&\times \lbrack
B_{ij}(p)+B_{ij}(k_{4})+B_{ij}(k_{3})-B_{ij}(k_{2})-B_{ij}(k_{1})]  \nonumber
\\
&&+\frac{E_{p}}{6N_{g}}\int_{1234}d\Gamma
_{123;4p}(1+f_{1}^{eq})(1+f_{2}^{eq})(1+f_{3}^{eq})f_{4}^{eq}(1+f_{p}^{eq})^{-1}
\nonumber \\
&&\times \lbrack
B_{ij}(p)+B_{ij}(k_{4})-B_{ij}(k_{3})-B_{ij}(k_{2})-B_{ij}(k_{1})].
\label{eq:constraint}
\end{eqnarray}
However, solving $B_{ij}$ using this equation is technically challenging. It
is easier to perform a projection (or convolution) to the above equation,
then solve for the less restricted $B_{ij}$. We will discuss the procedure
below.

By plugging Eq. (\ref{eq:constraint}) into Eq. (\ref{eq:eta}), we obtain $%
\eta $ in a bilinear form of $B_{ij}$, 
\begin{eqnarray}
\eta &=&\frac{1}{80T}\int \prod_{i=1}^{4}\frac{d^{3}k_{i}}{(2\pi )^{3}2E_{i}}
|M_{12\rightarrow 34}|^{2}(2\pi )^{4}\delta ^{4}(k_{1}+k_{2}-k_{3}-k_{4}) 
\nonumber \\
&&\times (1+f_{1}^{eq})(1+f_{2}^{eq})f_{3}^{eq}f_{4}^{eq}  \nonumber \\
&&\times \lbrack B_{ij}(k_{4})+B_{ij}(k_{3})-B_{ij}(k_{2})-B_{ij}(k_{1})]^{2}
\nonumber \\
&&+\frac{1}{120T}\int \prod_{i=1}^{5}\frac{d^{3}k_{i}}{(2\pi )^{3}2E_{i}}
|M_{12\rightarrow 345}|^{2}(2\pi )^{4}\delta
^{4}(k_{1}+k_{2}-k_{3}-k_{4}-k_{5})  \nonumber \\
&&\times (1+f_{1}^{eq})(1+f_{2}^{eq})f_{3}^{eq}f_{4}^{eq}f_{5}^{eq} 
\nonumber \\
&&\times \lbrack
B_{ij}(k_{5})+B_{ij}(k_{4})+B_{ij}(k_{3})-B_{ij}(k_{2})-B_{ij}(k_{1})]^{2}.
\label{eta_bi}
\end{eqnarray}
Then one can solve for $B_{ij}$ which equates the right hand sides of Eqs. (%
\ref{eq:eta}) and (\ref{eta_bi}). This is nothing but a projection of Eq. (%
\ref{eq:constraint}). The resulting solution is not unique because only the
projected equation but not the equation itself is satisfied. However, it is
proven \cite{Resibois,Arnold:2003zc} that the true solution of $B_{ij}$,
i.e. the solution satisfying Eq. (\ref{eq:constraint}), would give the
maximum value of $\eta $. Thus, solving for $\eta $ becomes a variational
problem.

Recently, an algorithm is developed to find the true solution of $B_{ij}$
systematically \cite{Chen:2010vg}. Thus, this approach is no more
variational but systematic. Here we outline the procedure. First, expanding $%
B(p)$ using a specific set of orthogonal polynomials \cite%
{Dobado:2001jf,Chen:2006iga}: 
\begin{equation}
B(p)=(E_{p}/T)^{y}\sum_{r=0}^{r_{\max }}b_{r}B^{(r)}(E_{p}/T),
\label{eq:expansion}
\end{equation}
with $y$ a constant chosen to be 1 in this case. The dimensionless
polynomial $B^{(r)}$ satisfying the orthonormal condition 
\begin{equation}
\int \frac{d^{3}p}{(2\pi )^{3}E_{p}}f_{p}^{eq}(1+f_{p}^{eq})|\mathbf{p}
|^{2}(E_{p}/T)^{y}B^{(r)}(E_{p}/T)B^{(s)}(E_{p}/T)=T^{4}\delta _{rs}.
\end{equation}
Then Eq. (\ref{eta_bi}) can be written in a compact form, 
\begin{equation}
\eta =\left\langle B\left\vert F\right\vert B\right\rangle
=\sum_{r,s=0}^{r_{\max }}b_{r}b_{s}\left\langle B^{(r)}\left\vert
F\right\vert B^{(s)}\right\rangle ,  \label{eq:eta-orth}
\end{equation}
while Eq. (\ref{eq:eta}) gives 
\begin{eqnarray}
\eta &=&\sum_{r=0}^{r_{\max }}b_{r}\frac{N_{g}}{15T}\int \frac{d^{3}p}{(2\pi
)^{3}E_{p}}f_{p}^{eq}(1+f_{p}^{eq})|\mathbf{p}
|^{2}(E_{p}/T)^{y}B^{(r)}(E_{p}/T)  \nonumber \\
&=&\sum_{r=0}^{r_{\max }}b_{r}L^{(r)}\delta _{r0}=b_{0}L^{(0)},
\label{eq:eta-orth1}
\end{eqnarray}
with 
\begin{equation}
L^{(0)}=\frac{N_{g}}{15T}B^{(0)}\int \frac{d^{3}p}{(2\pi )^{3}E_{p}}%
f_{p}^{eq}(1+f_{p}^{eq})|\mathbf{p}|^{2}(E_{p}/T)^{y}.
\end{equation}%
From Eqs. (\ref{eq:eta-orth}) and (\ref{eq:eta-orth1}), we can find $b_{0}$
by solving the equation 
\begin{equation}
L^{(r)}\delta _{r0}=\sum_{s=0}^{r_{\max }}b_{s}\left\langle
B^{(r)}\left\vert F\right\vert B^{(s)}\right\rangle ,
\end{equation}%
and then determine $\eta $ from Eq. (\ref{eq:eta-orth1}). [Note that there
could be more than one solution satisfying Eqs. (\ref{eq:eta-orth}) and (\ref%
{eq:eta-orth1}), but they all give the same $\eta $.]

In Ref. \cite{Chen:2010vg}, it is proven that this procedure gives a
monotonically increasing value of $\eta $ with increasing $r_{\max}$. Thus,
one can systematically approach the true value of $\eta $ by adding more
terms in the expansion of Eq. (\ref{eq:expansion}). We find good convergence
in this algorithm. From $r_{\max }=1$ to $2$, $\eta $ changes by less than $%
2\%$ for $\alpha _{s}\leq 0.3$. Better convergence is found for smaller $%
\alpha _{s}$.

\section{Numerical results}

\subsection{Leading-Log result}

The leading order [$O(\alpha _{s}^{-2})$] contribution to $\eta $ only
involves the 22 and 23 processes \cite{Arnold:2003zc}. The 22 collision rate
is larger than 23 by a $(\ln \alpha _{s})$ factor. In the leading-log (LL)
approximation, one just needs to focus on the small $q_{T}$ contribution
from the 22 process. Furthermore, it was shown in \cite%
{Baym:1990uj,Heiselberg:1994vy} that using the HTL regulator (\ref{eq:htl})
gives the same LL viscosity to that using the $m_{D}$ regulator (\ref{MD}).
Thus, after performing the small $q_{T}$ expansion to Eq. (\ref{eta_bi}), we
obtain 
\begin{equation}
\eta _{LL}\simeq 27.1\frac{T^{3}}{g^{4}\ln (1/g)},  \label{LL}
\end{equation}%
which coincides with that of \cite{Arnold:2000dr} to significant digits
shown above. Using the entropy density for non-interacting gluons, $s=N_{g}%
\frac{2\pi ^{2}}{45}T^{3}$ (for $m_{g}=0$), we obtain 
\begin{equation}
\frac{\eta _{LL}}{s}\simeq \frac{3.9}{g^{4}\ln (1/g)}.  \label{LLs}
\end{equation}%
This will be used to check our numerical result later.

\subsection{$\protect\eta _{22}$ -- Shear viscosity with the 22 process only}

\begin{figure}[tbp]
\caption{$\protect\eta _{22}$ over the entropy density (left panel) and $%
\protect\eta _{22}$ over $\protect\eta _{22(MD)}$ (right panel) in various
treatments. `LL' is the leading log result of Eq. (\protect\ref{LL}). `HTL'
is the result using the full HTL matrix element of Eq. (\protect\ref{eq:htl}%
). `MD' is the result using $m_{D}$ as the regulator as in Eq. (\protect\ref%
{MD}). `AMY' is AMY's result. }
\label{fig:1}\includegraphics[scale=0.4]{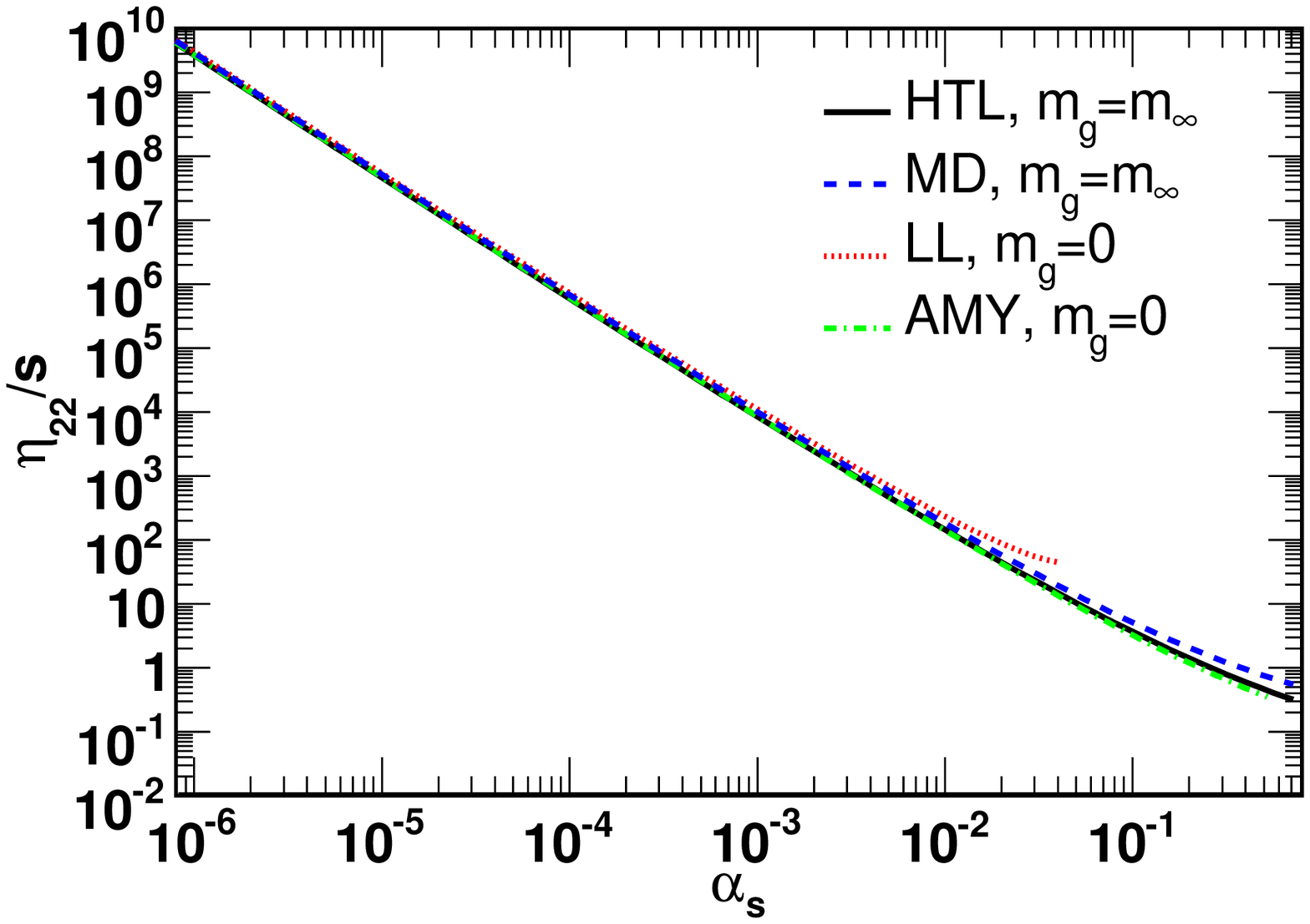} %
\includegraphics[scale=0.4]{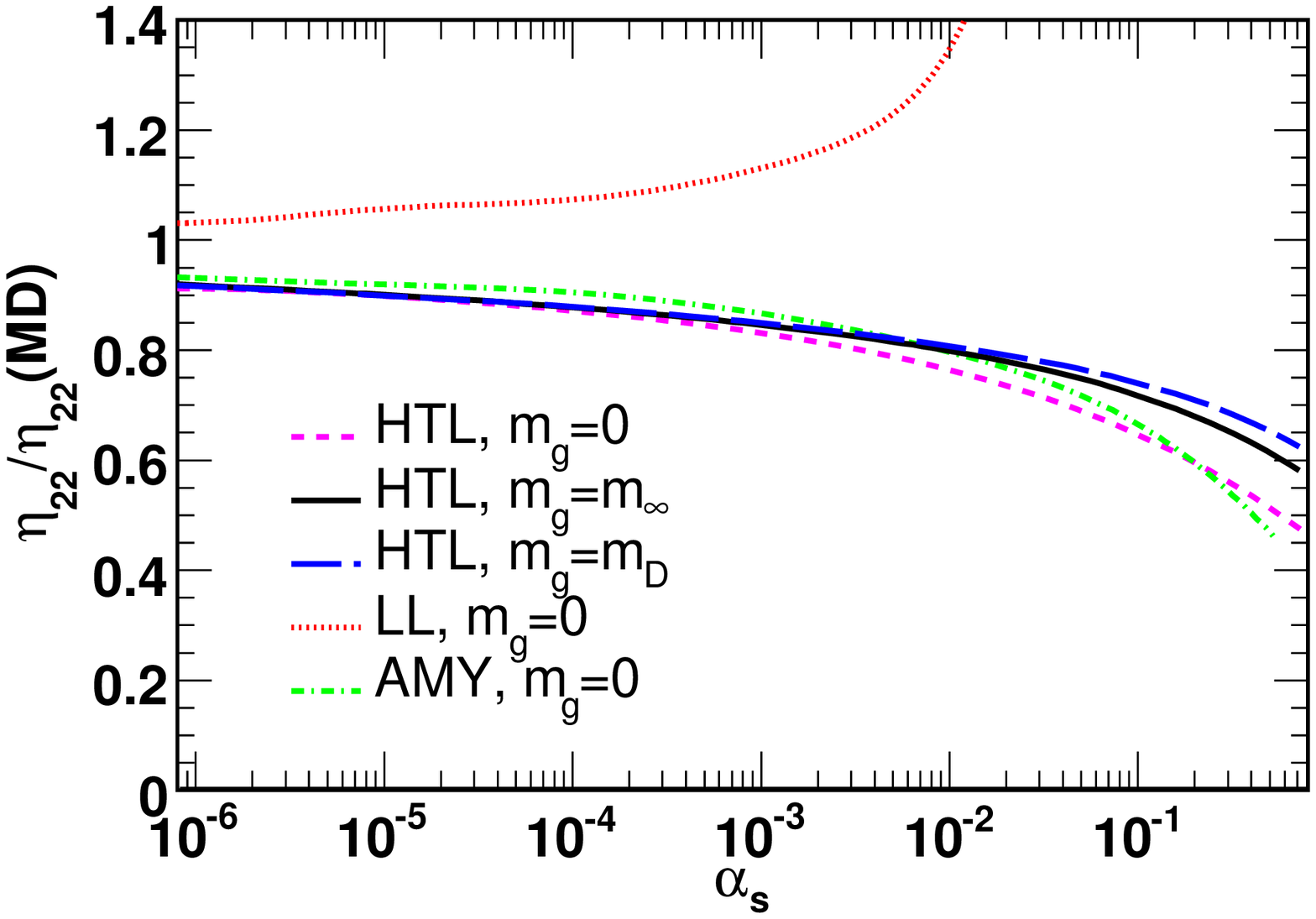}
\end{figure}

To study the effect of the HTL regulator, $\eta _{22}$ (i.e. $\eta $ with
the 22 process only) with the HTL and $m_{D}$ for the internal gluon masses,
respectively, are shown in Fig. \ref{fig:1}. The LL result $\eta _{LL}$ and
AMY's $\eta _{22}$ [denoted as $\eta _{22(AMY)}$] \cite%
{Arnold:2000dr,Arnold:2003zc} are also shown. The external gluon mass $m_{g}$%
, used in kinematics and in $f_{p}^{eq}$\ such that $E_{p}=\sqrt{\mathbf{p}%
^{2}+m_{g}^{2}}$, is a higher order effect in $\eta _{22}$. Changing $m_{g}$%
\ from 0 to $m_{D}=\sqrt{2}m_{\infty }$\ yields an $O(m_{\infty
}^{2}/T^{2})=O(\alpha _{s})$\ variation to $\eta _{22}$. This is confirmed
numerically in the left panel of Fig. \ref{fig:1}. It is a good check to our
numerical calculation that $\eta _{22(HTL)}$\ and $\eta _{22(MD)}$\ both
converge to $\eta _{LL}$\ in small $\alpha _{s}$, and $\eta _{22(HTL)}$\
agrees well with $\eta _{22(AMY)}$\ when $m_{g}=0$\ is used to conform with
the AMY result.

About the HTL effect, $\eta _{22(HTL)}/\eta _{22(MD)}$ is quite close to
unity at $\alpha _{s}=10^{-6}$, see the right panel of Fig. \ref{fig:1}.
This ratio gets smaller at larger $\alpha _{s}$ and reaches 0.65 at $\alpha
_{s}=0.1$ with little $m_{g}$ dependence (each $\eta _{22(HTL)}/\eta
_{22(MD)}$ is evaluated with the same $m_{g}$). This means the error is $%
\sim 30\%$ in the shear viscosity at $\alpha _{s}=0.1$ if we use $m_{D}$ as
the regulator for the gluon propagator instead of the HTL propagator.

\subsection{$\protect\eta _{22+23}$ -- Shear viscosity with the 22 and 23
processes}

In our full calculation, we use the HTL propagator for the 22 process.
However, for technical reasons, we use the internal gluon mass $m_{D}$ for
the 23 process. More specifically, we use matrix elements of Eqs. (\ref%
{eq:m3}-\ref{eq:htl},\ref{(ij)}), $E_{p}=\sqrt{\mathbf{p}^{2}+m_{g}^{2}}$ in
kinematics and $f_{p}^{eq}$ for external gluons, and $m_{g}=m_{\infty }$. If
the external gluons are massless but the internal gluons are massive, then
the $1/[(p_{1}+p_{2})^{2}-m_{D}^{2}]$ factor could diverge. Using $%
m_{g}=m_{\infty }$, each term in Eq. (\ref{eq:m3}) is non-negative. In AMY
and XG, external gluon masses were not included ($m_{g}=0$). This divergence
was avoided by keeping only the most singular matrix elements in the small $%
k_{\perp }$, $q_{\perp }$\ limit, and taking the collinear approximation
(AMY) or regulating the gluon bremsstrahlung infrared divergence by the
Landau-Pomeranchuk-Migdal (LPM) effect (XG) which will be discussed in
Section \ref{error}.

\begin{figure}[tbp]
\caption{Left panel: $\protect\eta _{22}/s$ and $\protect\eta _{22+23}/s$
for various cases. Right panel: the ratio of our result to AMY's.
`22(HTL)+23' denotes $\protect\eta _{22+23}$ where $m_{g}=m_{\infty }$, the
full HTL matrix element (\protect\ref{eq:htl}) is used for the 22 process,
and the matrix elements of Eqs. (\protect\ref{eq:m3}-\protect\ref{eq:htl},%
\protect\ref{(ij)}) are used for the 23 process. `22(AMY)+23(AMY)' denotes
AMY's result for $\protect\eta _{22+23}$ (we have used 
the same $s$ in $\eta _{22+23}/s$ as ours with $m_g=m_\infty$). 
The range of the `recommended value' of $\protect\eta _{22+23}$ is bounded by $\protect\eta %
_{-}$(lower bound) and $\protect\eta _{+}$(upper bound).}
\label{fig:2}\includegraphics[scale=0.4]{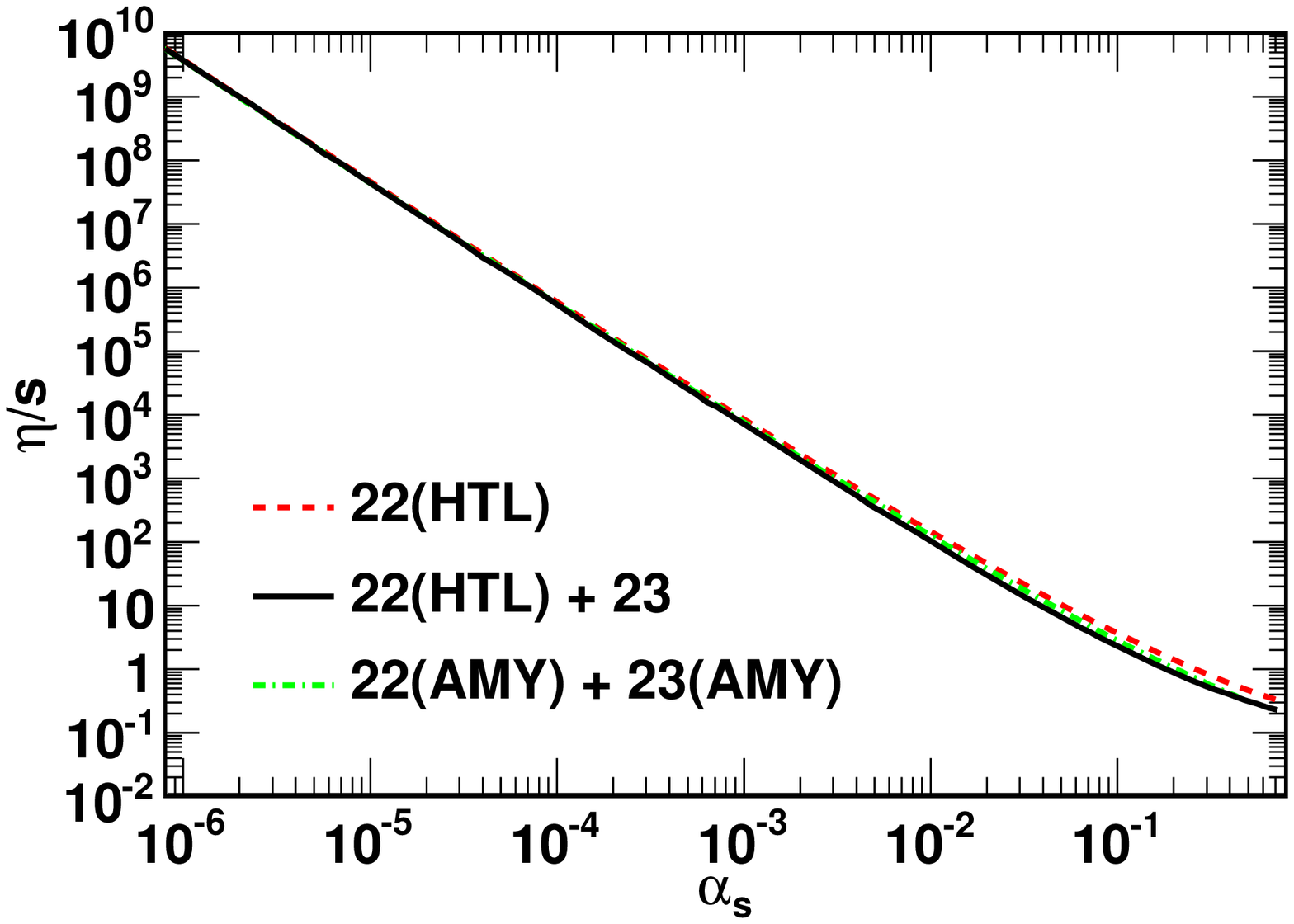} %
\includegraphics[scale=0.4]{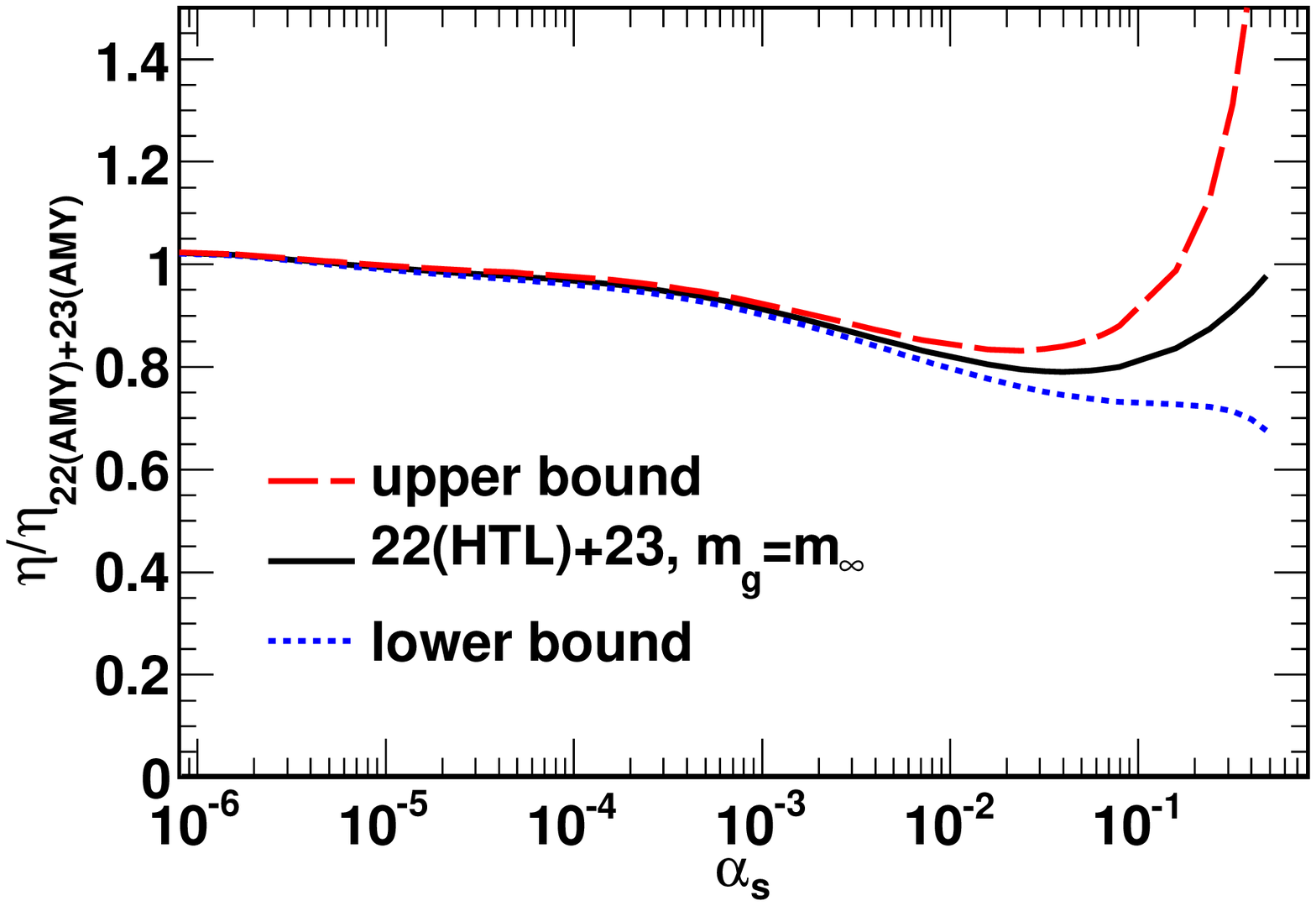}
\end{figure}

We show $\eta _{22}/s$ and $\eta _{22+23}/s$ in the left panel of Fig. 2,
where the HTL propagator is used for 22 and the external gluon mass 
$m_{g}=m_{\infty }$. We also show AMY's result for $\eta _{22+23}/s$ for
comparison. In the left panel, we see that our result for $\eta _{22+23}/s$ deviates
from XG's significantly, e.g. at $\alpha _s=0.1/0.01/0.001$ our result gives
2.3/103/7020 but XG give about 0.45/20/1200 (read off from 
Ref. \cite{Xu:2007ns}) respectively, i.e. our result is about 5 to 6 times 
as large as XG's. In Ref. \cite{Wesp:2011yy} they improve their calculation 
by using the Kubo relation and give larger values 0.795/60 at $\alpha _s=0.1/0.01$, 
so our result is about 2 to 3 times theirs. 
In the right panel of Fig. 2, we see that the ratio of our
result to AMY's approaches unity at $\alpha _{s}\lesssim 10^{-4}$ and $\sim
0.8$ at $\alpha _{s}\approx 0.004$. The deviation in moderate $\alpha _{s}$
is partly due to the finite angle, non-colinear 3-body configurations in the
23 process described by the full matrix element (\ref{eq:m3}) and partly due
to the gluon mass.  We have also included a theoretical error band for 
$\eta _{22+23}$ which will be discussed in Section \ref{error}.

\begin{figure}[tbp]
\caption{$\protect\eta _{22+23}/\protect\eta _{22}$ in various cases. `HTL'
and uses HTL gluon propagators for the 22 process. The range of the
`recommended value' of $\protect\eta _{22+23}$ is bounded by $\protect\eta %
_{-}$(lower bound) and $\protect\eta _{+}$(upper bound). The external gluon
mass is set to $m_{\infty }$. AMY's and XG's $\protect\eta _{22+23}/\protect%
\eta _{22}$ are also shown. XG's result is taken from Ref. \cite{Wesp:2011yy}.}
\label{fig:3}\includegraphics[scale=0.4]{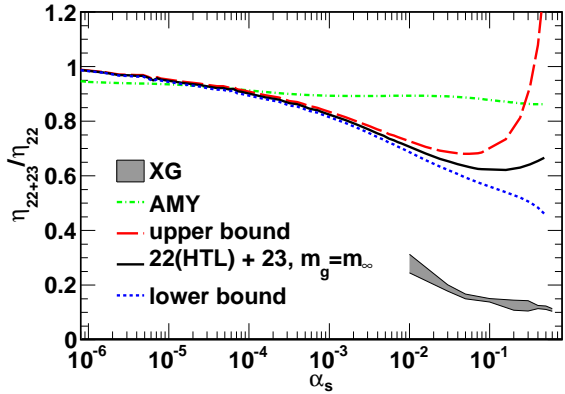}
\end{figure}

The effect of the 23 process can be seen more clearly in the ratio $\eta
_{22+23}/\eta _{22}$ shown in Fig. \ref{fig:3}. We have plotted $\eta
_{22(HTL)+23}/\eta _{22(HTL)}$ together with AMY's and XG's result for
comparison. Our result shows that the 22 process dominates at small $\alpha
_{s}$. When $\alpha _{s}$ increases, $\eta _{22+23}/\eta _{22}$ decreases
and the central value reaches the minimum of $\sim 0.6$ (which means the 23
collision rate is $\sim 60\%$ of the 22 one) at $\alpha _{s}\simeq 0.1$ and
then increases again for $\alpha _{s}\gtrsim 0.1$. Thus, our result shows:
(1) the 22 process is more important than 23 when $\alpha _{s}\lesssim 0.1$;
(2) the finite angle contributions are important at intermediate $\alpha
_{s} $\ ($\alpha _{s}\sim 0.01$-$0.1$). We see that AMY's result which
employs the collinear approximation for the $1\leftrightarrow 2$ process
(corresponding to our 23 process), gives $\eta _{22+23}/\eta _{22}$ close to
unity. This implies their 23 collisions is just a small purturbation to the
22 collisions. XG's result in Ref. \cite{Xu:2007ns} which employs the soft gluon bremsstrahlung
approximation, however, gives $\eta _{22+23}/\eta _{22}\simeq \lbrack
0.11,0.16]$ around 1/8. This implies their 23 collision rate is about 7
times the 22 one. Their improved result in Ref. \cite{Wesp:2011yy} 
gives $\eta _{22+23}/\eta _{22}\simeq [0.1,0.3]$ 
corresponding to the 23 collision rate that is about 2 to 9 times 
(on average 5 times) the 22 one. Our result takes neither of the approximations, 
lies between AMY's and XG's result, but it agrees with AMY's
result qualitatively but not with XG's.

\subsection{Error estimation}

\label{error}

\begin{table}[b]
\caption{$\protect\eta /s$ values for our $\protect\eta _{22(HTL)+23}/s$
result and the range of our `recommended values' bounded by $\protect\eta %
_{-}/s$ and $\protect\eta _{+}/s$.}
\label{tab}%
\begin{tabular}{|c|c|c|c||c|c|c|c|}
\hline
$\alpha _{s}$ & $\frac{\eta _{22(HTL)+23}}{s}$ & $\frac{\eta _{-}}{s}$ & $%
\frac{\eta _{+}}{s}$ & $\alpha _{s}$ & $\frac{\eta _{22(HTL)+23}}{s}$ & $%
\frac{\eta _{-}}{s}$ & $\frac{\eta _{+}}{s}$ \\ \hline
1.00E-6 & 3.75E+9 & 3.75E+9 & 3.75E+9 & 0.100 & 2.30 & 2.08 & 2.58 \\ \hline
1.58E-6 & 1.54E+9 & 1.54E+9 & 1.54E+9 & 0.125 & 1.67 & 1.48 & 1.91 \\ \hline
2.51E-6 & 6.33E+8 & 6.32E+8 & 6.34E+8 & 0.150 & 1.30 & 1.13 & 1.52 \\ \hline
3.98E-6 & 2.60E+8 & 2.60E+8 & 2.61E+8 & 0.175 & 1.06 & 0.909 & 1.27 \\ \hline
6.31E-6 & 1.07E+8 & 1.07E+8 & 1.08E+8 & 0.200 & 0.888 & 0.752 & 1.09 \\ 
\hline
1.00E-5 & 4.42E+7 & 4.40E+7 & 4.44E+7 & 0.225 & 0.765 & 0.638 & 0.971 \\ 
\hline
1.58E-5 & 1.83E+7 & 1.82E+7 & 1.84E+7 & 0.250 & 0.672 & 0.551 & 0.880 \\ 
\hline
2.51E-5 & 7.57E+6 & 7.52E+6 & 7.61E+6 & 0.275 & 0.600 & 0.484 & 0.811 \\ 
\hline
3.98E-5 & 3.14E+6 & 3.12E+6 & 3.16E+6 & 0.300 & 0.542 & 0.430 & 0.759 \\ 
\hline
6.31E-5 & 1.30E+6 & 1.29E+6 & 1.31E+6 & 0.325 & 0.495 & 0.386 & 0.720 \\ 
\hline
1.00E-4 & 5.43E+5 & 5.38E+5 & 5.47E+5 & 0.350 & 0.456 & 0.349 & 0.691 \\ 
\hline
1.58E-4 & 2.26E+5 & 2.24E+5 & 2.28E+5 & 0.375 & 0.423 & 0.318 & 0.669 \\ 
\hline
2.51E-4 & 9.47E+4 & 9.38E+4 & 9.55E+4 & 0.400 & 0.395 & 0.291 & 0.654 \\ 
\hline
3.98E-4 & 3.97E+4 & 3.93E+4 & 4.01E+4 & 0.425 & 0.371 & 0.268 & 0.644 \\ 
\hline
6.31E-4 & 1.67E+4 & 1.65E+4 & 1.68E+4 & 0.450 & 0.350 & 0.248 & 0.639 \\ 
\hline
1.00E-3 & 7.02E+3 & 6.94E+3 & 7.11E+3 & 0.475 & 0.332 & 0.230 & 0.638 \\ 
\hline
1.58E-3 & 2.97E+3 & 2.93E+3 & 3.01E+3 & 0.500 & 0.316 & 0.214 & 0.642 \\ 
\hline
2.51E-3 & 1.27E+3 & 1.25E+3 & 1.29E+3 & 0.525 & 0.302 & 0.200 & 0.649 \\ 
\hline
3.98E-3 & 542. & 532. & 553. & 0.550 & 0.289 & 0.187 & 0.659 \\ \hline
6.31E-3 & 234. & 229. & 240. & 0.575 & 0.278 & 0.175 & 0.673 \\ \hline
1.00E-2 & 103. & 99.9 & 106. & 0.600 & 0.267 & 0.164 & 0.691 \\ \hline
1.58E-2 & 45.6 & 44.1 & 47.3 & 0.625 & 0.258 & 0.153 & 0.711 \\ \hline
2.51E-2 & 20.7 & 19.8 & 21.7 & 0.650 & 0.250 & 0.144 & 0.736 \\ \hline
3.98E-2 & 9.63 & 9.08 & 10.2 & 0.675 & 0.242 & 0.135 & 0.764 \\ \hline
6.31E-2 & 4.62 & 4.28 & 5.01 & 0.700 & 0.235 & 0.127 & 0.795 \\ \hline
\end{tabular}%
\end{table}

Our $\eta _{22(HTL)+23}/s$ is tabulated in Table \ref{tab}. The error
assignment is based on the the following error analyses for $\eta _{22+23}$:

(a) HTL corrections for the 23 process. From our $\eta _{22}$ error
analysis, we assign a $\sim 30\%$ error at $\alpha _{s}=0.1$ to the 23
contribution for not implementing the HTL approach to the 23 collisions. The
error will be smaller at smaller $\alpha _{s}$ if the scaling for $\eta
_{22} $ holds also for the error. Since the HTL effect tends to reduce the
magnetic screening effect which lowers the IR cut-off and enhances the 23
collision rate, the HTL correction tends to reduce $\eta _{22+23}$.

(b) LPM effect. We will try to estimate the error from neglecting the LPM
effect. An intuitive explanation of this effect was given in Ref. \cite%
{Gyulassy:1991xb}: for the soft bremsstrahlung gluon with transverse
momentum $k_{T}$, the mother gluon has a transverse momentum uncertainty $%
\sim k_{T}$ and a size uncertainty $\sim 1/k_{T}$. It takes the
bremsstrahlung gluon the formation time $t\sim 1/\left( k_{T}v_{T}\right)
\sim E_{k}/k_{T}{}^{2}$ to fly far enough from the mother gluon to be
resolved as a radiation. But if the formation time is longer than the mean
free path $l_{mfp}\approx O(\alpha _{s}^{-1})$, then the radiation is
incomplete and it would be resolved as $gg\rightarrow gg$ instead of $%
gg\rightarrow ggg$. Thus, the resolution scale is set by $t\leq l_{mfp}$.
This yields an IR cut-off $k_{T}^{2}\geq E_{k}/l_{mfp}\approx O(\alpha _{s})$
on the phase space \cite{Wang:1994fx}. Thus, the LPM effect reduces the 23
collision rate and will increase $\eta _{22+23}$. Our previous calculation
using the Gunion-Bertsch formula shows that implementing the $m_{D}$\
regulator gives a very close result to the LPM effect \cite{Chen:2009sm}.
Thus, we will estimate the size of the LPM effect by increasing the external
gluon mass $m_{g}$ from $m_{\infty }$ to $m_{D}$.

(c) Higher order effect. The higher order effect is parametrically
suppressed by $O(\alpha _{s})$, but the size is unknown. Computing this
effect requires a treatment beyond the Boltzmann equation \cite%
{Hidaka:2010gh} and the inclusion of the 33 and 24 processes. We just
estimate the effect to be $\alpha _{s}$ times the leading order which is $%
\sim 10\%$ at $\alpha _{s}=0.1$.

Combining the above analyses, we consider errors from (a) to (c). 
To compute a recommended range for $\eta /s$, we will work with the $R_{22}$
and $R_{23}$ collision rates defined as 
\begin{eqnarray}
R_{22}^{-1} &\equiv &\eta _{22(HTL)},  \nonumber \\
\left( R_{22}+R_{23}\right) ^{-1} &\equiv &\eta _{22(HTL)+23}.
\end{eqnarray}%
Using HTL instead of $m_{D}$ in for the gluon propagator enhances the 22
rate by a factor of 
\begin{equation}
\delta \equiv \frac{R_{22(HTL)}}{R_{22(MD)}}=\frac{\eta _{22(MD)}}{\eta
_{22(HTL)}}.
\end{equation}%
We will assume that the same enhancement factor appears in 23 rate as well,
such that 
\begin{equation}
\delta \simeq \frac{R_{23(HTL)}}{R_{23(MD)}}.
\end{equation}%
On the other hand, the LPM effect is estimated to suppress the 23 rate by a
factor of 
\begin{equation}
\gamma =\frac{R_{23}\left( m_{g}=m_{D}\right) }{R_{23}\left( m_{g}=m_{\infty
}\right) }.
\end{equation}

Combining the estimated HTL and LPM corrections to the 23 rate, the 22+23
rate is likely to lie in the range $[R_{22}+R_{23},R_{22}+\gamma \delta
R_{23}]$, while the higher order effect gives $\pm \alpha _{s}\left(
R_{22}+R_{23}\right) $ corrections to the rate. Without further information,
the errors are assumed to be Gaussian and uncorrelated, the total rate is 
\begin{equation}
\left( R_{22}+\frac{\gamma \delta +1}{2}R_{23}\right) \pm \left( \frac{%
\gamma \delta -1}{2}R_{23}\right) \pm \alpha _{s}\left( R_{22}+R_{23}\right)
,
\end{equation}
and the recommended upper ($\eta _{+}$) and lower ($\eta _{-}$) range for $%
\eta _{22+23}$ are 
\begin{equation}
\eta _{\pm }=\frac{1}{\left( R_{22}+\frac{\gamma \delta +1}{2}R_{23}\right)
\mp \sqrt{\left( \frac{\gamma \delta -1}{2}R_{23}\right) ^{2}+\alpha
_{s}^{2}\left( R_{22}+R_{23}\right) ^{2}}}.
\end{equation}
The $\eta _{\pm }$ values are shown in the right panel of Fig. \ref{fig:2}
and in Fig. \ref{fig:3}.

Our final $\eta /s$ result is as presented in Table I. At $\alpha _{s}=0.1$, 
$\eta /s\simeq \left[ 2.08, 2.58 \right] $, which is closer to $2.85$ of AMY
than to XG's $0.5$ in Ref. \cite{Xu:2007ns} and $0.795\pm 0.025$ in Ref. \cite{Wesp:2011yy}. 
At $\alpha _{s}=0.3$, we have $\eta /s\simeq \left[ 0.43, 0.759 \right] $, 
which is compatible to $0.6$ of AMY but not to XG's $0.13$ in Ref. \cite{Xu:2007ns} 
and $0.166\pm 0.025$ in Ref. \cite{Wesp:2011yy}.

\subsection{When does the $\protect\alpha _{s}$ perturbation break down?}

We have only carried out the leading order [$O(\alpha _{s}^{-2})$] $\eta /s$
in the $\alpha _{s}$ expansion. Without computing the higher order
contribution, it is hard to tell at what value of $\alpha _{s}$ the
perturbation starts to break down. In the above section, we have naively
assumed the higher order contribution to be the leading order times $\alpha
_{s}$, i.e. the expansion breaks down at $\alpha _{s}\simeq 1$. However,
explicit computations of thermal dynamical quantities and transport
coefficients showed that the break down might happen at smaller $\alpha _{s}$
\cite{Blaizot:2000fc,Andersen:2002ey,CaronHuot:2007gq} (the screening mass
computation breaks down at $\alpha _{s}\simeq 0.1$ \cite{Andersen:2002ey}
and the heavy quark diffusion constant computation breaks down at $\alpha
_{s}\simeq 0.01$\cite{CaronHuot:2007gq}).

Looking more closely to our leading order $\eta /s$ shown in the left panel
of Fig. 2, the $\alpha _{s}$ dependence of $\eta /s$ changes qualitatively
at $\alpha _{s}\simeq 0.1$. This could be a sign that higher order $\alpha
_{s}$ dependence has become as important as the leading order one. Thus, the
higher order corrections could be bigger than our previous estimation, and
our result might be only reliable when $\alpha _{s}\lesssim 0.1$.

Having said that, it is interesting that our $\eta /s$ bends slightly upward
at $\alpha _{s}\gtrsim 0.1$ as if $\eta /s$ is trying to avoid going below
the conjectured $1/4\pi $ bound. Several models have proposed to described
the microscopic picture in the non-perturbative region \cite%
{Asakawa:2006tc,Hidaka:2009ma,Khvorostukhin:2010cw,Bluhm:2010zi}. And a
similar result to ours is obtained in a recent calculation \cite{Das:2010yi}
based on one kind of simplification of the 23 matrix element \cite%
{Das:2010hs}.

Our result implies that the proposed perfect fluid limit $\eta /s\simeq
1/(4\pi )$ cannot be achieved by perturbative QCD alone.

\section{Conclusions}

We have calculated the leading order [$O(\alpha _{s}^{-2})$] $\eta /s$ of a
gluon plasma in perturbation using the kinetic theory. The leading order
contribution only involves the 22 and 23 processes. The leading order
contribution only involves the 22 and 23 processes. The Hard-Thermal-Loop
(HTL) propagator has been used for the 22 matrix element, while the exact
matrix element in vacuum is supplemented by the Debye mass $m_{D}$ for gluon
propagators for the 23 process. Also, the asymptotic mass $m_{\infty }=m_{D}/%
\sqrt{2}$ is used for the external gluon mass in the kinetic theory as well.
The errors from not implementing HTL and the Landau-Pomeranchuk-Migdal
effect in the 23 process, and from the uncalculated higher order [$O(\alpha
_{s}^{-1})$] corrections, have been estimated.

Our result for $\eta /s$ lies between that of Arnold, Moore and Yaffe (AMY)
and Xu and Greiner (XG).\ Our result shows that although the finite angle
contributions are important at intermediate $\alpha _{s}$\ ($\alpha _{s}\sim
0.01$-$0.1$), the 22 process is still more important than 23 when $\alpha
_{s}\lesssim 0.1$. This is in qualitative agreement with AMY's result. We
find no indication that the proposed perfect fluid limit $\eta /s\simeq
1/(4\pi )$\ can be achieved by perturbative QCD alone.

\vspace{0.3cm}

Acknowledgement: We thank G. Moore for providing us their tables of $\eta /s$%
. JWC and QW thank KITPC, Beijing, for hospitality. JWC is supported by the
NSC and NCTS of ROC. QW is supported in part by the `100 talents' project of
Chinese Academy of Sciences and by the National Natural Science Foundation
of China under grant 10735040. HD is supported in part by the Natural
Science Foundation of the Shandong province under grant ZR2010AQ008. JD is
supported in part by the Innovation Foundation of Shandong University under
grant 2010GN031.

\end{document}